# From Prompting to Partnering: Personalization Features for Human-LLM Interactions

SI THU

School of Computer Science, Faculty of Engineering & IT, University of Technology Sydney, sithu@alumni.uts.edu.au

A. BAKI KOCABALLI

School of Computer Science, Faculty of Engineering & IT, University of Technology Sydney, baki.kocaballi@uts.edu.au

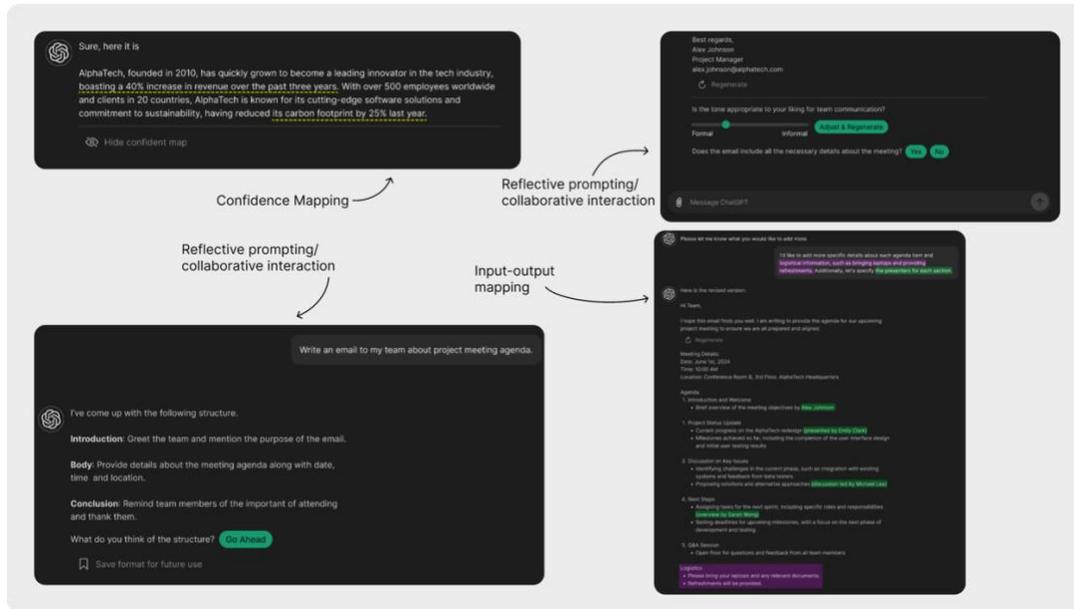

Figure 1. Personalisation for improving human-LLM interactions

**Abstract**. Large Language Models (LLMs), such as ChatGPT, exhibit advanced capabilities in generating text, images, and videos. However, their effective use remains constrained by challenges in prompt formulation, personalization, and opaque decision-making processes. To investigate these challenges and identify design opportunities, we conducted a two-phase qualitative study. In Phase 1, we performed in-depth interviews with eight everyday LLM users after they engaged in structured tasks using ChatGPT across both familiar and unfamiliar domains. Our findings revealed key user difficulties in constructing effective prompts, iteratively refining AI-generated responses, and assessing response reliability especially in domains beyond users' expertise. Informed by these insights, we designed a high-fidelity prototype incorporating Reflective Prompting, Section Regeneration, Input-Output Mapping, Confidence Indicators, and a Customization Panel. In Phase 2, user testing of the prototype indicated that these interface-level improvements may prove useful for reducing cognitive load, increasing transparency, and fostering more intuitive and collaborative human-AI interactions. Our study contributes to the growing discourse on human-centred AI, advocating for human-LLM interactions that enhance user agency, transparency, and co-creative interaction, ultimately supporting more intuitive, accessible, and trustworthy generative AI systems.

**CCS CONCEPTS** • Human-centered computing → Human computer interaction (HCI) → Empirical studies in HCI

**Additional Keywords and Phrases:** AI personalization, ChatGPT, Google Gemini, User Profiling, Generative AI customization, Figma prototype usability

## 1 INTRODUCTION

The rapid adoption of generative AI tools like ChatGPT surpassing 100 million users in just two months [2] signals a paradigm shift in how humans interact with technology. Among these technologies, Large Language Models (LLMs) excel at generating coherent text, answering complex questions, and aiding creative tasks [3, 4]. Yet, their transformative potential remains unevenly accessible: while experts adeptly navigate prompt engineering to refine outputs, novices face steep learning curves, generic responses, and opaque decision-making processes [5]. For instance, similar prompts can yield contradictory results, and users often lack the domain knowledge to diagnose failures [5, 27, 42]. LLMs today appear to prioritize technical capability over human-centered interaction, leaving non-experts frustrated and disengaged.

Prior work highlights two interrelated challenges. First, **prompt engineering**, a skill critical for eliciting desired outputs remains inaccessible to non-experts [60]. Studies show that even minor wording changes can yield vastly different results [48], and users often lack the domain knowledge or iterative refinement skills to navigate this generative variability (non-deterministic outputs) [9, 27], often abandoning interactions after repetitive corrections [14]. Second, **personalization**, tailoring interactions to individual needs is still underexplored in LLMs. While advances like LaMP [32] and Persona-DB [33] demonstrate the technical feasibility of user-adaptive models, most LLMs today rely on static prompts rather than leveraging user history, intent, or contextual preferences [30]. This omission leads to generic outputs, inefficiency, and frustration, particularly for novices [6]. Compounding these issues are ethical risks: personalization requires sensitive data, raising privacy concerns [25], while opaque systems risk amplifying biases or misinformation [7, 8].

To address this gap, this work explores features designed to reduce the cognitive burden of prompt engineering by focusing on two key areas: interface-level personalization and enhanced collaborative interaction models. Interface-level personalization involves implementing features at the graphical user interface (GUI) level—such as prompts, controls, and visualizations—rather than making algorithmic or model-level changes. This approach typically requires less sensitive personal data, making it more practical and user-friendly. On the other hand, enhanced collaborative interaction models address the limitations of current one-directional interactions between users and LLMs. At present, users issue commands, but LLMs neither clarify user intent (e.g., by asking clarifying questions) nor provide transparent justifications for their outputs (e.g., through confidence scores) [9]. This "black box" dynamic can stifle shared understanding and trust.

Through interviews with eight everyday ChatGPT users and subsequent evaluation of a high-fidelity prototype, our study provides empirical insights into: i) user struggles with prompt engineering and desires for tailored interactions, and ii) how explainability features and customization controls through a more collaboration interaction model enhance user experience

## 2 BACKGROUND

### 2.1 User experience in LLM interactions

A critical barrier to LLM adoption is the expertise asymmetry in prompt engineering. While experts leverage nuanced prompting strategies to achieve desired outputs [12], novices struggle with ambiguity in phrasing, context provision, and iterative refinement [43]. Users face high metacognitive demands, including self-monitoring of prompt clarity, decomposing complex tasks, and evaluating the reliability of outputs processes that often lead to frustration and abandonment [39]. Trust is further eroded by misalignment between user expectations and model behaviour; for example, Dhillon et al. [28] found that even when LLMs improve output quality, users report diminished satisfaction due to perceived loss of ownership and control. These challenges highlight the need for interfaces that mitigate cognitive load while fostering transparency.

## 2.2 Prompt engineering challenges and support

Prompt engineering plays a critical role for generating effective LLM outputs, yet it presents significant challenges. Korzyński et al. [60] identify it as a key digital skill, calling for more research into its empirical, ethical, and evaluative dimensions. White et al. [59] provide structured approaches for crafting prompts, while Song et al. [46] highlight how LLMs' non-deterministic nature adds variability, necessitating adaptive, context-aware practices [38, 51]. Steyvers et al. [29] emphasize the gap between user confidence in LLM outputs and the models' internal confidence, advocating tailored explanations to align expectations and build trust. Ethical concerns are also prominent: Vinay et al. [26] show how emotionally charged prompts can amplify disinformation risks, and Chen et al. [12] call for standardized evaluative frameworks to assess prompt design's impact on AI behaviour and user experience [6, 27]. To address these issues, Shin et al. [11] propose automated methods for generating prompts, reducing reliance on user expertise and increasing accessibility to LLM technologies [5]. Collectively, these studies stress the need for a balance between technical advancements and ethical accountability in prompt engineering development.

Effective prompt engineering support balances AI assistance with user autonomy. Dhillon et al. [28] showed that scaffolding such as next-sentence suggestions can improve output quality without overwhelming users, while Weisz et al. [27] proposed design principles like "Design for Co-Creation" and "Design for Generative Variability" to manage generative variability. However, existing tools fall short in providing real-time guidance, context retention, or intuitive controls for non-experts. For example, Vinay et al. [26] revealed that emotionally charged prompts significantly influence output quality, yet novices lack frameworks to apply such strategies. Similarly, Li et al. [31] identified a demand for more visual and interactive prompt design tools, particularly for users without technical backgrounds.

## 2.3 Personalization in LLM Interfaces

Beyond making prompt engineering more intuitive, personalization has emerged as a key factor for improving user engagement and satisfaction with generative AI. Wang et al. [30] emphasize how adaptable models responding to individual user preferences can streamline LLM-based interactions. Similarly, Salemi et al. introduce the LaMP benchmark, integrating user profiles into LLM evaluations to achieve more tailored outputs [32]. Sun et al. [33] propose the Persona-DB framework, enhancing LLM personalization with higher accuracy and reduced retrieval size, while Qin et al. [34] introduce a privacy-preserving on-device framework for user-specific content generation. Richardson et al. [35] suggest a summary-augmented approach combining retrieval techniques with LLM-generated user summaries to tackle cold-start problems and runtime constraints. Tankelevitch et al. [39] further argue that personalization can reduce cognitive load by aligning system behaviour with user mental models, but this potential remains underexplored in GUI design.

## 3 METHODS

This study employed a qualitative approach to understand and enhance user experiences with LLMs, focusing on OpenAI's ChatGPT. In the first phase, in-depth interviews were conducted with eight everyday LLM users to explore their backgrounds, familiarity with AI, and use cases. Participants interacted with ChatGPT to answer their in-domain and out-of-domain questions (i.e., questions within or out of their area of expertise and knowledge), with the option to ask up to two follow-up questions per type. At the time of conducting the interviews, ChatGPT-4o had not been released. Consequently, seven out of eight participants used Version-3.5 to complete the activity, while only one participant (P1) used Version-4. Based on insights from the interviews, a high-fidelity prototype was developed in Figma (Figures 1 & Figure 2), incorporating personalized features such as frequently asked prompts, adaptive suggestions, and customization options.

In the second phase, six participants from the first phase tested the prototype by completing guided tasks and providing feedback. Following the development of the Figma design prototype, participants from the initial interview phase were invited to evaluate the prototype. The testing process was conducted in two stages: task completion and feedback collection. During the task completion phase, participants were guided through a structured activity using the prototype. For the feedback collection phase, participants were asked to reflect on their overall experience with the design. Given that the prototype was a high-fidelity representation of the ChatGPT interface rather than a fully functional system, participants were provided with detailed instructions on how to navigate each screen. The task required participants to instruct ChatGPT to compose an email regarding a project meeting for their team. This scenario was selected based on interview insights into common use cases and was divided into multiple steps to showcase the integration of all five proposed features (see Table 1).

Feedback was gathered through multiple channels to ensure a comprehensive understanding of participants' experiences. Observations were made during the task phase, noting participants' reactions, expressions, and spontaneous comments. Following the interaction, participants were interviewed about their experience to elicit additional insights. All data was then transcribed. Thematic analysis was conducted after both phases to uncover key themes from the interview data and prototype feedback. Two independent coders performed the analysis, resolving differences through consensus.

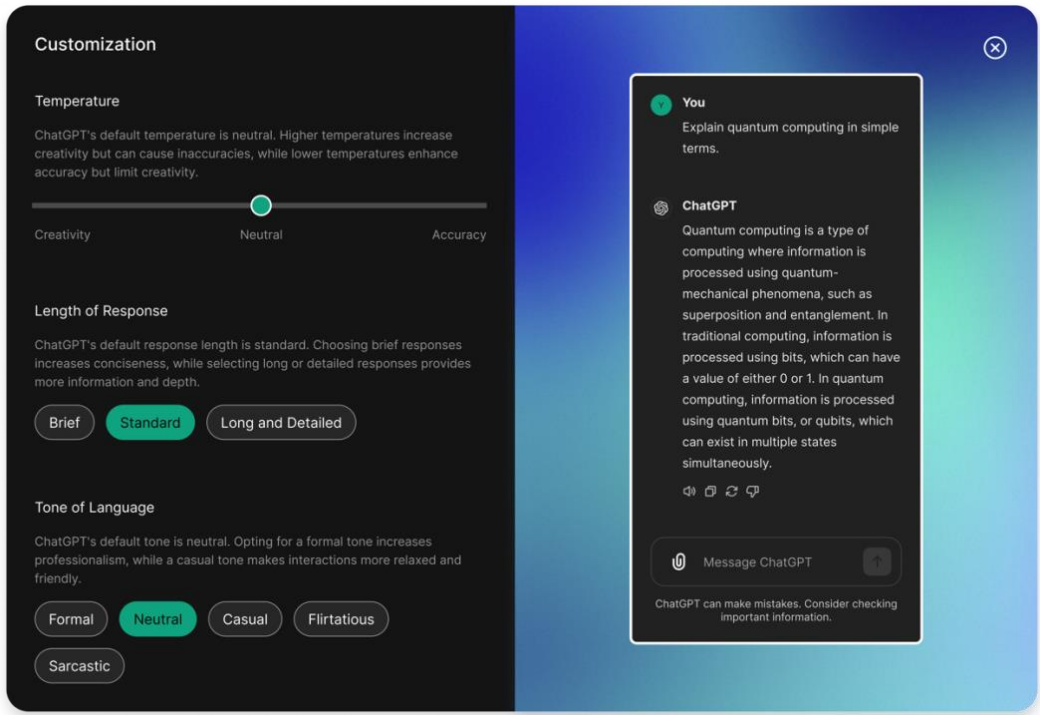

**Figure 2**. **Customization panel**

## 4 RESULTS

### 4.1 Study 1 – Everyday LLM use and experience

*4.1.1 Participants' Experience and Familiarity*

Participants exhibited a spectrum of familiarity with generative AI, ranging from novice to advanced users. Advanced users (e.g., P1) integrated ChatGPT into daily workflows for tasks like coding and email drafting, while intermediate users (P2–P8) leveraged it selectively for brainstorming and information retrieval. A common trajectory emerged: initial experimentation with "fun" queries (e.g., P4: "*I started with funny questions, but now it's part of my life*") evolved into structured use for domain-specific tasks. However, even advanced users like P1 faced challenges in prompt refinement, suggesting a steep learning curve in mastering AI interactions.

*4.1.2 Prompting Experiences and Response Quality*

Across participants, prompt engineering emerged as both a challenge and a critical skill for maximizing ChatGPT's usefulness. While most participants began with simple, direct queries, they discovered that iterative refinement and explicit context setting led to significantly more accurate, in-depth, and contextually aligned responses. P7 characterized this process succinctly: "*Sometimes I have to refine my prompts multiple times to get the answer I'm looking for.*" This iterative approach required users to break down complex requests into smaller sub-questions or provide structured directives (e.g., "act as a [role]"). Participants also learned to rephrase or add constraints to prompts. For example, P4, who frequently asked coding-related questions, noted that specifying programming language details or requesting certain code structures (e.g., iterative vs. recursive) helped reduce irrelevant or too-general content.

Despite these strategies, the quality and specificity of ChatGPT's responses varied with domain familiarity. When participants asked about topics in which they had expertise (e.g., ERP systems for P1 or programming tasks for P4), they felt better equipped to recognize and correct inaccuracies. In these scenarios, ChatGPT was viewed as highly reliable and efficient. However, when tackling unfamiliar subjects—for instance, P6 asking about aerospace engineering or P2 exploring state-of-the-art AI tools—participants reported more frequent uncertainties or generic content. This challenge was particularly pronounced when participants had little background knowledge to assess the responses' accuracy. As P6 stated: "*It looks really, really comprehensive… but I can't evaluate if it's actually accurate.*" Hallucinations or overly broad answers occasionally surfaced, causing participants to second-guess ChatGPT's reliability. Some participants also noticed superficial domain coverage, where ChatGPT provided factually plausible but contextually shallow explanations. They remedied this by engaging in a multi-turn dialogue, prompting for more depth or clarifying contradictory points. Although this iterative prompting often improved the final output, participants highlighted time costs and cognitive load as drawbacks.

*4.1.3 Personalization and Tailored Responses*

A recurring theme in participants' reflections was the desire for personalization, with many envisioning an AI assistant that could remember user-specific preferences and context across multiple interactions. Despite ChatGPT's ability to retain information within a single conversation thread, most participants found that responses remained largely generic or one-size-fits-all, particularly when they were seeking nuanced or context-sensitive advice (e.g., personalized workout programs or domain-specific protocols). P1 highlighted the lack of continuity between sessions, noting that preferences expressed in one conversation were not carried over into another. Similarly, P7, who explored health and fitness recommendations, wished that ChatGPT could automatically adapt to personal goals and constraints—rather than having to restate them in every new prompt: "*I wish ChatGPT could remember my preferences and provide more personalized responses based on my previous interactions.*"

To achieve meaningful personalization, participants overwhelmingly supported the idea of user profiles. This mechanism could allow users to input or store details—such as profession, skill level, dietary restrictions, or learning goals—that ChatGPT could use to tailor its responses. Several participants believed that profile-based adaptation would shorten the iterative prompting process, as the AI would already have baseline context. P3 reiterated how useful this could be: "*If I could add my profile details and ChatGPT could use that to give me personalized responses, that would be amazing.*" Finally, participants voiced frustration that ChatGPT sometimes assumed knowledge they did not possess or, conversely, explained content at too low a level. The ability to fine-tune responses to users' prior knowledge—akin to adaptive learning systems—was frequently cited as a beneficial next step.

These findings highlight critical gaps in LLM interfaces that must be addressed to improve user experience. **Reducing Cognitive Load** is essential, as users struggle with iterative prompt refinement; dynamic scaffolding, such as adaptive auto-complete and role-based templates, can simplify this process. **Building Trust** is another priority, as users often question the accuracy and reasoning behind AI outputs; incorporating explainability features like input-output mapping and confidence indicators can enhance transparency. Finally, **Enabling Co-Creation** requires moving beyond transactional interactions to foster personalized, adaptive, context-aware dialogues, where AI proactively engages users with tailored suggestions and iterative feedback, creating a more collaborative and intuitive experience.

**4.2 Prototype design features**

Based on the key findings of the first phase, we developed a high-fidelity prototype with five features in Figma, incorporating a style guide to replicate ChatGPT's interface (Table 1):

| Design Feature | Description | Phase 1 Findings |
| --- | --- | --- |
| **1. Reflective Prompting** | Offers real-time, context-aware completions based on interaction patterns to streamline prompt creation. | Reducing Cognitive Load |
| **2. Section Regenerate[1]** | Engages users in iterative dialogue (clarifying questions, multiple "regenerate" buttons) to refine outputs collaboratively. | Enabling Co-Creation |
| **3. Input-Output Mapping** **4. Confidence Indicator** | Visualizes the influence of input on output and confidence indicator to highlight lower-confidence sections. | Building Trust |
| **5. Customization Panel** | Allows users to adjust parameters (temperature, response length, tone) to align outputs with intent. | Building Trust & Enabling Co-Creation |

**Table 1**. High-fidelity prototype features mapped to the Phase 1 findings

**4.3 Study 2 – Prototype evaluation**

Six participants (P1, P2, P3, P4, P6, P7) from the Phase 1 agreed to evaluate the Figma prototype through a guided task of composing an email, demonstrating its features. Feedback was collected via direct questions and observations, capturing verbal and non-verbal reactions, with P2 providing continuous feedback during the session. This section present participants' perceptions of five key features (see Figures 1 and 2) in our LLM prototype interface–Reflective Prompting, Section Regeneration, Input-Output Mapping, Confidence Mapping, and the Customization Panel–based on feedback from six participants. We note recurring themes, highlight illustrative quotes, and summarize design implications.

*4.3.1 Reflective Prompting*

**Supportive but unfamiliar**. Across participants, Reflective Prompting elicited both curiosity and occasional confusion. Several participants noted its potential to reduce cognitive load. For instance, P3 stated, "Like it because it reduces the load on the user", and P1 appreciated how "*the process was supported by ChatGPT itself, so it takes a lot of thinking away*." However, unfamiliarity with receiving reflective questions from an AI was a recurring theme. While participants acknowledged reflective prompts' potential to surface deeper thinking, P2 worried about having "*too many reflective-prompt questions*." Participants' comments indicate a desire for more **selective engagement**, providing controls to tailor the frequency or depth of prompts. Participants also suggested refining the visual and interactive aspects. P1 requested

---

[1] This feature was introduced to ChatGPT's Canvas mode in early October 2024.

"more button options" for context, and P3 proposed clearer separation between AI-generated content and reflective questions (e.g., "*Question and buttons can have better visual separation from the response content*"). From a usability standpoint, these findings highlight the need for i) finer-grained customization of the reflective prompts' content or context, and ii) explicit onboarding and visual cues to help users recognize and trust an unfamiliar AI-driven feature.

*4.3.2 Section Regeneration*

**Granular editing of outputs.** Section Regeneration (SR) appealed to participants who desired fine-grained editing controls with P2 noting the benefit of having "*many SR buttons*". Participants highlighted the usefulness of regenerating only specific sections instead of the entire response, reducing repetition and better targeting the portion in need of refinement. However, they indicated that this feature needed clearer labelling and contextual help. P4 suggested using more specific signifiers such as "*Regenerate Introduction*," while P1 requested "*more information*" upon tapping regenerate (e.g., clarifying what changes might occur). Together, these insights suggest that users may benefit from more **transparency** supported by explicit, segment-specific labels, and previews of expected edits.

*4.3.3 Input-Output Mapping and Confidence Indicator*

**Improved traceability and time savings**. Most participants described Input-Output Mapping as one of their favourite features. P2 called it "*intuitive*" and their "*Favorite feature*," while P6 referred to it as a "*game changer*." By allowing users to visually track how specific parts of their input map to corresponding outputs, the feature "*saves time from reading the whole response again*" (P5) and helps users "*understand what is being addressed*" (P1). Participants liked that they could more easily compare incremental changes, with P3 pointing out it is "*hard to track changes throughout the process*" otherwise. Such positive reactions align with prior work on explainability and traceability in AI systems: making the model's interpretive process more visible fosters user trust and comprehension [57].

**Source transparency.** The visual indicators, the yellow underlines, were perceived as novel and trust-enhancing. P1 and P6 immediately understood the mapping, with P1 noting it "*enhances my trust*." However, participants sought richer explanations of confidence levels. P3 and P5 proposed adding percentage scales, while P6 recommended proactive underlining (without toggling) and in-tab fact-checking links. A recurring theme was the desire to know "*the source of ChatGPT*" or "*where responses come from*" (P1, P3, P6). Participants sought more transparency about how the underlying model arrived at particular outputs. Several participants requested side-by-side layouts (P3, P5) to further ease comparison of input segments and outputs; they felt scrolling between prompt and answer to see color-coded highlights was not optimal (P5). Suggestions included "*better colour choice*" (P5) and expanded explanation tooltips.

*4.3.4 Customisation Panel*

**Multi-level customisation and more granular response settings**. The Customization Panel received positive remarks for enabling adjustments like tone, style, or length. P6 noted it "*makes it fun to use,"* while P2 praised the "*tone-of-language slider*" for post-hoc adjustments, and P5 appreciated being able to "*tweak the tone of a response once it has been generated,*" reinforcing its dynamic nature. Participants noted that they could reduce iterative prompting by manipulating sliders and toggles in the panel (P6). Several participants (P1, P2, P3) wanted customization settings specific to each "tab," "chatroom," or "type of conversation." For instance, P1 advocated "*more customization for each tab or chat room*," while P2 did not want a one-size-fits-all default. P3 further suggested advanced features like adjusting "*length of response*" near the input box, highlighting a desire for more granular or context-dependent controls. Finally, participants requested clarity on how the CP interacts with direct user instructions in prompts (P1) and recommended additional layers of customization, such as using "infographics" (P6) or "language level" (P3).

# 5 DISCUSSION

## 5.1 Summary of key findings

Our two-phase study highlights both the promise and the complexity of using Large Language Models (LLMs) like ChatGPT in everyday contexts. In Phase 1, participants reported steep learning curves in prompt engineering, frequently struggling to generate context-rich queries and resorting to iterative refinement to improve the AI's responses. They expressed a consistent desire for more personalized interactions—aiming to reduce the repetitive need to restate user preferences and background knowledge. Phase 2, centred on evaluating our high-fidelity prototype, reinforced these findings: participants appreciated features that lowered cognitive load (e.g., Reflective Prompting), improved trust through transparency (e.g., Input-Output Mapping and Confidence Indicators), and enabled more collaborative user-AI exchanges (e.g., Section Regenerate, Customization Panel). Personalization emerged as a key factor in enhancing user experience, with participants expressing a strong desire for tailored responses. This aligns with existing literature, which underscores the role of personalization in boosting engagement and satisfaction [32, 33]. Participants showed willingness to share some personal data for improved interactions, though the prototype lacked the backend functionality to implement such features. Previous research confirms that personalization in LLMs can enhance accuracy and efficiency while safeguarding user privacy [21, 22, 23, 24]. Overall, our findings highlight three core challenges for LLM-based systems: (1) reducing cognitive burden in prompt formulation and correction; (2) building trust through transparent AI processes; and (3) enabling adaptive, collaborative interactions that cater to personal contexts.

## 5.2 Design implications

### 5.2.1 Reducing Cognitive Load

A major challenge across participants was formulating prompts that adequately convey context and constraints, echoing similar observations by Korzyński et al. [60] and Weisz et al. [27]. Reflective Prompting helped users break down complex requests, while Section Regenerate avoided regenerating entire texts. This "just-in-time" support resonates with Dhillon et al. [28], who note that contextual scaffolding eases iterative refining. However, preventing over-scaffolding is crucial to preserve user autonomy [28]. Providing customization controls, letting users set the depth or frequency of reflective prompts may mitigate this tension, ensuring that users can fine-tune the level of AI assistance to their own preferences or expertise.

### 5.2.2 Fostering Trust through Explainability

Trust emerged as a multifaceted construct in our study, influenced by participants' ability to understand how an LLM interprets prompts and generates outputs. The prototype's Input-Output Mapping underlined key segments of user input to show their corresponding impact on the AI's response, reducing the sense of a "black box", a practice supported by prior work emphasizing the importance of "explanatory debugging" [64]. Meanwhile, Confidence Indicators addressed user concerns regarding correctness by highlighting sections of the AI's output that might require additional scrutiny, resonating with Steyvers et al. [29] and Chen et al. [12] who advocate for aligning user and model confidence. These features collectively support "explainable AI" principles, reinforcing research that ties system usability and trust to transparency [27, 21]. Yet, our results reveal a persistent user desire for deeper insight. Participants wanted to know the rationale behind certain suggestions or the specific sources that informed them. Designers should thus consider layered explanations offering both a high-level overview (e.g., color-coded highlights) and more granular details (e.g., a tooltip with confidence scores or reference links). This tiered approach enables users with different knowledge levels to access explanations at varying depths, consistent with best practices in HCI for progressive disclosure [65].

*5.2.3 Facilitating Collaborative Interaction*

Participants expressed enthusiasm for a more reciprocal relationship with the LLM. Moving beyond single-shot queries, our prototype introduced collaborative features such as Section Regenerate (enabling targeted edits) and a Customization Panel (allowing adjustments to tone, style, and length). These mechanisms reframed the AI interaction from a transactional exchange into an iterative, co-creative process where users felt an increased sense of ownership, reflecting the "Design for Co-Creation" principle put forth by Weisz et al. [27] and the adaptive design recommended by Salemi et al. [32] in user-personalized LLM scenarios. Ultimately, personalization drives co-creation: participants mentioned that if the LLM "remembered" their context such as professional roles, personal communication style, or ongoing project details, it could anticipate user needs, thus reducing repetitive re-prompting. However, personalization raises additional considerations (e.g., privacy and data governance) also highlighted by Qin et al. [34] and Vinay et al. [26], underscoring the importance of designing ethically responsible systems that store user data securely and transparently.

## 5.3 Limitations

Our study has some limitations with its focus on a relatively small sample of active or potential LLM users, which may limit the generalizability of findings to more diverse user populations. Furthermore, the prototype was tested as a simulation rather than a fully functional system. Future research could extend to longitudinal studies with larger and more diverse user groups and incorporate real-time usage analytics. Additionally, examining domain-specific LLM applications (e.g., healthcare, and education) could uncover nuanced requirements and ethical constraints beyond general-purpose usage.

## 6 CONCLUSION

This study investigated the challenges everyday users face interacting with LLMs and evaluated interface-level improvements designed to promote personalization, transparency, and collaboration. In Phase 1, participants interacting with ChatGPT reported high cognitive load from iterative prompt refinement, mistrust due to opaque processes, and dissatisfaction with generic AI responses. To mitigate these challenges, Phase 2 introduced and evaluated a novel prototype incorporating five interface improvements: Reflective Prompting, Section Regeneration, Input-Output Mapping, Confidence Indicators, and Customization Panels. Preliminary findings suggest these interface-level improvements hold promise in reducing cognitive load, enhancing transparency, and promoting more intuitive human-AI collaboration. However, the study's findings are constrained by a small sample size and reliance on a simulated prototype. Future research should explore longitudinal user experiences, evaluate fully operational prototypes in authentic settings, and investigate domain-specific interactions to deepen understanding and refine these personalized features further.